\documentclass[aps,reprint,groupedaddress,longbibliography,superscriptaddress]{revtex4-1}
\usepackage{graphicx, tabularx, longtable, nicefrac, amsmath,enumerate, supertabular, eurosym}
\usepackage{hyperref,xcolor}
\hypersetup{colorlinks = true, citecolor = blue!30!black, linkcolor = blue!30!black}
\usepackage{booktabs}
  \DeclareUnicodeCharacter{2003}{}

\begin{document}
\title[Studying physics during the COVID-19 pandemic]{Studying physics during the COVID-19 pandemic: Student assessments of learning achievement, perceived effectiveness of online recitations, and online laboratories }

\author{P. Klein}\email{pascal.klein@uni-goettingen.de}
\affiliation{Faculty of Physics, Physics Education Research, University of Göttingen, Friedrich-Hund-Platz, 37077 Göttingen, Germany}

\author{L. Ivanjek}
\affiliation{Faculty of Physics, Physics Education Research, Technische Universit\"at Dresden, Haeckelstraße 3, 01069, Dresden, Germany}

\author{M.N. Dahlkemper}
\affiliation{Faculty of Physics, Physics Education Research, University of Göttingen, Friedrich-Hund-Platz, 37077 Göttingen, Germany}

\author{K. Jeličić}
\affiliation{Department of Physics, Faculty of Science, University of Zagreb, Bijenička cesta 32, 10000, Zagreb, Croatia}

\author{M.-A. Geyer}
\affiliation{Faculty of Physics, Physics Education Research, Technische Universit\"at Dresden, Haeckelstraße 3, 01069, Dresden, Germany}

\author{S. Küchemann}
\affiliation{Department of Physics, Physics Education Research Group, Technische Universit\"at Kaiserslautern, Erwin-Schr\"odinger-Str. 46, 67663 Kaiserslautern, Germany}

\author{A. Susac}
\affiliation{Faculty of Electrical Engineering and Computing, University of Zagreb, Unska 3, 10000 Zagreb, Croatia}

\date{\today}

\begin{abstract}
The COVID-19 pandemic has significantly affected the education system worldwide that was responding with a sudden shift to distance learning. While successful distance teaching requires careful thinking, planning and development of technological and human resources, there was no time for preparation in the current situation caused by COVID-19. Various physics courses such as lectures, tutorials, and the laboratories had to be transferred into online formats rapidly, resulting in a variety of simultaneous, asynchronous, and mixed activities. To investigate how physics students perceived the sudden shift to online learning, we developed a questionnaire and gathered data from N = 578 physics students from five universities in Germany, Austria, and Croatia. In this article, we report how the problem-solving sessions (recitations) and laboratories were adapted, how students’ judge different formats of the courses and how useful and effective they perceive them. The results are correlated to the students’ self-efficacy ratings and other behavioral measures (such as self-regulated learning skills) and demographics (such as the availability of quite learning spaces). In a related article, we focus on the online physics lectures and compare simultaneous vs. asynchronous teaching and learning methods [reference to be put in after acceptance]. This study is descriptive in nature and a survey study design was implemented to examine the relationships among variables.  We find that good communication abilities ($r=0.50$, $p<0.001$) and self-organization skills ($r=0.63$, $p<0.00$1) are positively correlated for perceived learning achievement. Furthermore, the previous duration of studies had a significant impact on the students’ perceived overall learning achievement [$F(5, 567)=13.1$, $p<0.001$], on the students’ acquisition of experimental skills during the physics laboratories [$F(3, 204)=4.41$, $p<0.01$], and on their assessment of the recitations' effectiveness $[F (4, 389) = 5.04$, $p= 0.001$]. That is, students in their first academic year show consistently lower scores than more progressed students. For the physics laboratories, it was found that gathering real data was crucial to the acquisition of experimental skills $F(2, 198) = 14.2$, $p<0.001$] and the reinforcement of content [$F(2, 216)=3.50$, $p=0.03$]. For the physics recitations, handing in own solutions for feedback was correlated with perceived effectiveness [$F(1, 393) = 6.74$, $p=0.01$]. We draw conclusions and implications for future online classes in which online activities will definitely take part since no return to exclusive presence teaching can be expected.

\end{abstract}

\maketitle

\section{Introduction}
The COVID-19 pandemic has significantly affected the education system worldwide that was responding with a sudden shift to distance learning. Although some forms of online teaching and learning existed in the system, the whole education community was suddenly forced in spring 2020 into an unplanned and unwanted remote teaching. Distance teaching requires careful thinking, planning, and development of technological and human resources for the successful accomplishment of expected learning outcomes. However, in the current situation caused by the COVID-19 pandemic, there was very little time for preparation; the instructors had to act quickly and to adapt to remote teaching. In the process, they had support from their organizations (schools and universities) in providing e-learning platforms and other digital learning management systems and communication tools. Nevertheless, the main burden was on the instructors to adjust their teaching methods and materials to an online format. The question rises whether their teaching approach is still efficient when taken from the physical classroom and transferred to technological devices. The aim of this study was to evaluate teaching and learning processes during the COVID-19 pandemic in the context of physics courses at the university level. This article focuses on the physics problem-solving sessions (so-called recitations) and the physics laboratories. It is structured as follows: First, a short overview about online teaching and learning in the context of physics is provided, particularly focusing on recitations (\ref{sec:recitations}), laboratories (\ref{sec:laboratories}), and influencing factors (\ref{sec:influ}) that were considered important for the study. In Sect. \ref{methods}, we describe the methods of data collection, questionnaire development and participating universities (samples), followed by the results, their discussion and conclusion.

Many students were taking part in online courses before the COVID-19 pandemic and in some countries, like the United States, distance education enrollments steadily increased in the last two decades  \cite{Seaman18}. New learning platforms, such as the massive open online courses (MOOCs), offer a new model for instruction and initial analysis has shown that student improvement is similar to the traditional face-to-face courses but significantly lower than interactive face-to-face courses  \cite{Colvin14}. Online education transforms all components of teaching and learning, and numerous educational researchers are trying to evaluate the results of these changes (e.g., \cite{Kebritchi17,Shapiro17,Davis18}). Online education processes differ from traditional face-to-face teaching in some features, so specific evaluation elements have to be introduced such as evaluation of the technology  \cite{Benigno2000} or different types of interaction treatments  \cite{Bernard09}. Previous physics education research (PER) studies have shown that some online teaching materials can be beneficial for students and could probably successfully substitute some aspects of face-to-face instruction. For example, multimedia learning modules using text, narration, flash animations, etc. had a positive effect on student performance on conceptual tests and classroom discussions  \cite{Sadaghiani11, Sadaghiani12}. Further, Kestin \textit{et al.} (2020) \cite{Kestin20} have shown that online lecture demonstrations helped students to learn more than traditional live demonstrations because they were more likely to make correct observations from the video. Positive experience with flipped classrooms in which content is delivered before the lecture, including online discussion, might also be informative and indicate a possible direction for online teaching during the current COVID-19 pandemic situation \cite{Wood16, Miller16}. Recently, Morphew \textit{et al.} (2020) \cite{Morphew20} found that students benefited from problem solution videos and this positive effect did not depend on the style of solution video. However, not all online resources were proved to be helpful for students. For example, a study on online instructional videos in an introductory mechanics course found that students used  laboratory videos which helped them complete the laboratories  more often than lecture videos, and there was little correlation between student engagement with the videos and their performance in the course \cite{Lin17}. Videos were supplemental course material in this study, so additional studies are needed to evaluate the benefit of instructional videos in online physics courses. These research results can provide guidance to instructors on how to produce effective online teaching materials but the question of how to engage students in the learning process remains.

\subsection{Recitations and homework problems} \label{sec:recitations}
Probably the most widespread forms of online tools in on-campus physics courses before the COVID-19 pandemic were Web-based homework systems, such as LON-CAPA, Mastering Physics, Expert TA, WebAssign, etc. which are usually incorporated in typical introductory physics courses, especially in the United States (they are not so common in Europe). They usually contain a large number of textbook problems with hints for students and provide automatic grading. In early studies, the online homework systems were evaluated and compared to ungraded homework and standard methods of collecting and grading homework  \cite{Bonham01, Cheng04}. Most studies have shown that online homework systems are beneficial  \cite{Thoennessen96,Demirci07, Kortemeyer08, Evans17} but usually more iterations and evolution of the homework system are needed to achieve the required efficiency  \cite{Gutmann18}. A recent review on the benefits of online versus traditional homework on students' performance showed that in half of the studies there were no differences in student performance for two homework formats whereas, in less than a third of the studies, the students' performance was higher in the online compared with the traditional homework format \cite{Magalhaes20}.  The online homework systems described above cover only one part of physics courses and could not be used as a replacement for face-to-face teaching or recitations during the COVID-19 pandemic. 

A major part of the university physics studies in Germany and most European countries consists of problem-solving sessions, so-called recitations. They provide an opportunity for students to work together in a small group on challenging problems that were designed to build conceptual understanding and problem-solving skills. Students apply critical-thinking skills and problem-solving methods rather than repeating the material covered in the textbooks or lectures. Two forms of recitations can be distinguished; (1) live sessions where students work together on the problems under the supervision of an instructor/lecturer or (2) working on the problems in spare time on a weekly basis (typically also in groups), submitting the solutions and receiving grading and feedback. In both cases, guided discussions take place after students engaged with the material.

\subsection{Physics laboratories} \label{sec:laboratories}
Physics laboratories are an important part of many physics courses or are special independent courses. Besides traditional physics labs, several types of online physics labs can be described: video analysis of instructor supplied videos of experimental procedures, virtual labs, hands-on student experimentation performed off-site, or remote labs  \cite{Reagan12}. In the last decade, we are witnessing a growing use of novel physics lab learning environments that are designed by using platforms, such as Arduino  \cite{Kubinova15, Kuan16, LaRocca20} and Raspberry Pi  \cite{Mandanici20}. With the proliferation of smartphones, the use of smartphone sensors in performing physics school experiments has also increased  \cite{Vogt12, Sukariasih19, Staaks}. Remote laboratories allow conducting real (laboratory) experiments remotely \cite{Groeber} whereas virtual laboratories are based on computer simulations, such as PhET  \cite{Finkelstein05}, which enable essential functions of laboratory experiments. Comparisons of physical and virtual laboratories in physics education showed advantages for each type of laboratory, so some researchers suggest combining the two to strengthen the learning process  \cite{deJong13, Brinson15, Husnaini19}. A recent review found overall positive effects of remote labs on cognitive, behavioral, and affective learning outcomes although the authors concluded that evaluation approaches of the learning outcomes in the reviewed studies were quite superficial  \cite{Post19}. An eye-tracking study has shown learning differences in virtual (simulation-based) and physical (microcomputer-based) laboratories. The results indicated that in simulation-based laboratories students tend to start experimenting, and then think about the questions on the worksheets, whereas for physical laboratories, they tend to think before doing  \cite{Chien15}. Since the current situation with online courses was not anticipated, new virtual and remote labs could not be established. The  online labs during the pandemic can be described as  a continued instruction of  courses that were  originally considered as lab courses prior to the transition to online teaching. In many cases, students were supported with (artificial or real) data sheets and a video of the experiment. Within this study, we paid particular attention to the evaluation of physics laboratories conducted online or on campus during the COVID-19 semester and we pay attention to the types of data that students engaged with.

\subsection{Influencing factors of online teaching and learning} \label{sec:influ}

Since many instructors did not have experience with organizing computer-based teaching, different practices of engaging students in online learning can be informative. For example, intelligent tutoring systems (i.e., systems to provide immediate and customized instruction or feedback to learners without human teachers) have been used in physics education for more than two decades \cite{Reif99, vanLehn05}, and an early review has shown that intelligent tutoring systems can be nearly as effective as human tutoring  \cite{vanLehn11}. Early intelligent tutoring systems were often developed for only a few topics, such as Newton’s laws  \cite{Reif99} or problems in the context of pulleys  \cite{Myneni13}, and were not in wide use. Recent advances in the design of web applications have enabled the development of new generations of tutoring systems  \cite{Schroeder15, Gladding15, Nakamura16, Ryan16, Heckler16, Mikula17} for helping students in developing physics problem-solving skills. 

Recent studies have reported the challenges of engaging all students in self-paced interactive electronic learning tutorials  \cite{Devore17, Marshman20}. The authors concluded that "many students in need of out-of-class remediation via self-paced learning tools may have difficulty motivating themselves and may lack the self-regulation and time management skills to engage effectively with tools specially designed to help them learn at their own pace."  \cite{Devore17} and they showed that students who worked in a supervised manner performed significantly better than students working on their own  \cite{Marshman20}. Self-regulatory, motivational, and social characteristics of students play an important role in their engagement with peers in online forums \cite{Traxler18} and online physics courses in general  \cite{Pond17}. This might be an important issue for learning processes during the pandemic, so we decided to evaluate several behavioral measures of students' learning. 

First, students' \textit{self-organization abilities} in general and during the COVID-19 semester was taken into account. We measured the degree to which students are metacognitively, motivationally, and behaviorally active participants in their own learning process, i.e. to which extend students can take a proactive role in monitoring their learning, maintaining motivation, and engaging in behaviors (e.g. study strategies) that lead to academic success.

Second, we also aimed at evaluating students' \textit{communication} (with their peers and instructors). Certainly, the influence of  interactions on knowledge construction is so ubiquitous that a  proper  understanding of learning cannot be  achieved without taking into account its social dimension. However, it can be assumed that it was easiest in the rapid transition to remote learning to have students work individually. Since learning at universities is done within a  social context, it therefore  becomes important to understand how communication between the lecturer and students, and among students, is perceived in an (almost exclusively) online learning environment.

Third, the \textit{technical conditions}, e.g. the accessibility to a computer and stable internet connection as well as their environment were considered, e.g. whether students had access to a quiet learning place for attending video conferences. 

Fourth, students'\textit{attitudes towards online learning} were in the scope of our research to identify their general attitude towards online learning.

Last, we asked for students'  perceived \textit{learning achievement and expected learning outcomes}. Students' beliefs about their capabilities or likelihood of success reveal important information about different course structures,  might vary between different groups of students, and presumably is correlated to the factors listed above.

\subsection{Research questions}
Since the physics courses evaluated in this study were not planned as online courses, standard procedures to assess the learning process in distance teaching could not be simply applied to the current remote teaching during the COVID-19 pandemic. So, the first step in the study was to develop a questionnaire for the assessment of different elements of the physics courses (lectures, laboratories, problem-solving sessions/recitations). This was done through discussions of university physics instructors who are also PER researchers and based on previous research findings and analysis of 20 semi-structured interviews with students on their experience with physics courses during the pandemic. The questionnaire was administered to 578 physics students from five universities in Germany, Austria, and Croatia. In this article, we will report the results on the effect of various behavioral aspects on perceived learning effectiveness during the COVID-19 term in spring/summer 2020 and the evaluation of online problem-solving sessions and physics laboratories. Other results will be reported in a related article (ref). Since the questionnaire was administered only once, reconsidering item formulations, adding/removing items based on an iterative process will follow. This study is descriptive in nature. A survey study design was implemented to examine the relationships among variables. However, we believe that priority is to report the initial data now since the planning for future terms affected by the COVID-19 pandemic is ongoing. We hope that the analysis of the collected data will provide an insight into the effects of the pandemic on the teaching and learning processes within university physics courses and will help instructors to improve their online courses during the current COVID-19 pandemics and for future online courses in general.

In this paper, the following four research questions are addressed:
\begin{enumerate}
\item[RQ1] What is the relationship between perceived learning effectiveness during the COVID-19 summer term and the various behavioral aspects that are important in digital teaching?
\item[RQ2] What are the differences between students of different levels of study progress concerning the behavioral aspects and learning effectiveness?
\item[RQ3] What formats were used to establish online problem-solving sessions and how have they been rated by the students in terms of effectiveness?
\item[RQ4] What formats were used to establish online physics laboratories and how have they been rated by the students in terms of effectiveness?
\end{enumerate}

While the first and second research questions  address all physics students that participated in the study, the third and fourth question are specific to the students who participated in the corresponding courses, i.e., problem-solving sessions and physics laboratories, respectively. The effectiveness was assessed in terms of students self-reports and not by performance tests or examinations due to the design of the study as a large-scale anonymous mid-end-term assessment. 

\section{Methods} \label{methods}
\subsection{Questionnaire development}
Based on common literature for evaluating online teaching and learning, the authors identified several aspects that were considered relevant for physics students’ learning in the current situation of the compulsory and involuntary change to online teaching. The aspects included evaluation of simultaneous and asynchronous activities, students’ attitudes towards online learning, communication abilities, technical and social aspects, and self-organization abilities. Furthermore, course structures that were typical for physics studies were also in the scope of interest, viz. the recitations, the physics laboratories and the physics labs for prospective teachers. Semi-structured interviews were then conducted with 20 students, and the transcripts were used to formulate the items of the questionnaire in an iterative process during close exchange between all researchers included in the study.  
The final questionnaire included 246 technical data fields and is available in English, German, and Croatian. The English versions of the questionnaire is presented in the appendix. While a rigorous analysis and validation process of the instrument is not in the scope of this paper, we restrict the presentation of instrument characteristics to the measures presented in Sect. 4.1. In the work presented here, we focus on the students' attitudes towards online learning, communication abilities, technical and social aspects, and self-organization abilities as well as the recitations and physics laboratories. For the latter two, the organization of the courses has also been assessed, e.g. whether students gathered data on their own (labs) or were provided, or whether the exercise sheets were solved live in online meetings or as homework followed by online discussions. Evaluation of the physics lectures and seminars in simultaneous and asynchronous formats are part of another work (reference)
\begin{table*}[t!]
\centering
\caption{Information about the sample}
\label{Tab:sample}

\begin{tabular}{lcccccccc} \hline\hline 
	&		&		&&	\multicolumn{5}{c}{Duration of studying physics (Years of physics studied)}	 \\ 
	&	Total sample	&	Male (\%)	&&	~~$<$ 1 yr~~	&	~~1--2.5 yrs~~	&	~~3--4.5 yrs~~	&	~~5--6 yrs~~	&	~~$>$ 6 yrs~~	\\ \hline \addlinespace[3pt]
Dresden	&	114	&	73 (64.0)	&&	39 (34.2)	&	30 (26.3)	&	30 (26.3)	&	12(10.5)	&	1 (0.9)	\\
Göttingen	&	232	&	144 (62.1)	&&	83 (35.8)	&	62 (26.7)	&	62 (26.7)	&	19 (8.2)	&	6 (2.6)	\\
Kaiserslautern	&	9	&	5 (55.6)	&&	4 (44.4)	&	2 (22.2)	&	…	&	2 (22.2)	&	1 (11.1)	\\
Wien	&	138	&	85 (61.6)	&&	30 (21.7)	&	65 (47.1)	&	24 (17.4)	&	11 (8.0)	&	7 (5.1)	\\
Zagreb	&	85	&	45 (52.9)	&&	25 (29.4)	&	27 (31.8)	&	23 (27.1)	&	6 (7.1)	&	4 (4.7)	\\ \addlinespace[4pt]
total	&	578	&	352 (60.9)	&&	181 (31.3)	&	186 (32.2)	&	139 (24.2)	&	50 (8.7)	&	19 (3.3)	\\\hline \hline

\end{tabular}

\end{table*}
\subsection{Participating universities}
\textit{University of Göttingen}. The University of Göttingen is a public research university located in Göttingen, Germany. It is the oldest university in the state of Lower Saxony and the largest in student enrollment, which stands at around 31,600. For studying physics, the faculty offers several degree programs, i.e., the Bachelor of Science in Physics, the Master of Science in Physics (Course language: English), the two subjects Bachelor with Physics (usually taken by teacher students), and the Master of Education with Physics.

\textit{TU Dresden}. The Technische Universit"at Dresden is a public research university, the largest university in Saxony and one of the 10 largest universities in Germany with 32,000 students. It is one of the oldest colleges of technology in Germany, and one of the country's oldest universities. For studying physics, the faculty offers several degree programs, the Bachelor of Science in Physics, the Master of Science in physics, and Teaching Studies.

\textit{University of Vienna}. The University of Vienna is the oldest university in the German-speaking world, located in Vienna, Austria, with an enrollment of more than 90,000 students.  The faculty of physics offers B. Sc., M. Sc., and teacher student programs.  

\textit{University of Zagreb}. The University of Zagreb is the oldest and the largest public university in Croatia. Typically, more than 65 000 students are enrolled. For studying physics, the Faculty of Science offers six programs of the integral undergraduate and graduate study of physics (Master of Science in Physics, Master of Education in Physics, Master of Education in Mathematics and Physics, Master of Education in Physics and Computer Science, Master of Education in Physics and Chemistry, and Master of Education in Physics and Technology).

\textit{TU Kaiserslautern}. The Technische Universität Kaiserslautern is a research university in Kaiserslautern, Germany. Approximately 15,000 students are enrolled at the moment. The physics department offers several degree programs, e.g.  Bachelor of Science in Physics, biophysics, physical engineering, and Master of Science.

\subsection{Data collection and sample}
The questionnaire was imported into Questback and distributed to physics students via mailing lists in mid-June and late July. This is therefore a mid-end term evaluation for all participating universities.  If certain course formats were not taken (e.g. the physics labs or the recitations), no questions were asked.
The distribution of the questionnaire was done either centrally by the faculties’ deans of studies or by faculty members. 

\begin{figure}[h!]
\includegraphics[width=.8\linewidth]{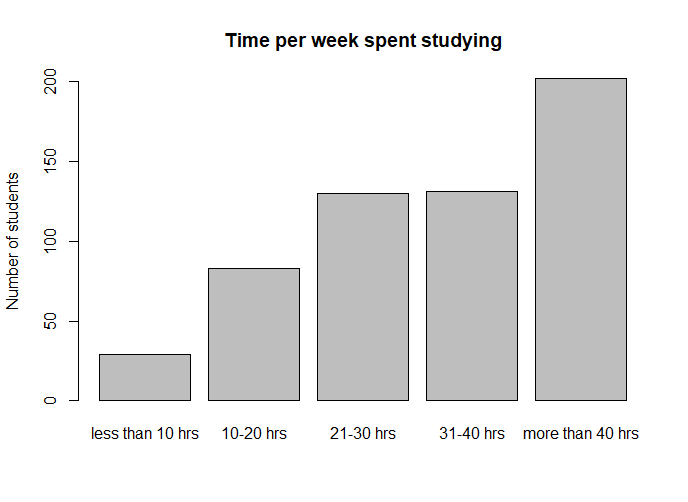}\\
\caption{Time per week spent studying}
\label{fig:sample}
\end{figure}

In total, 578 physics students (352 male) completed the questionnaire. The average time for answering the questionnaire was 24 min 37 sec.  Table \ref{Tab:sample} shows the distribution of the locations. Most students who participated in the study were in early semesters (63.5\% of the sample studied for less than 2.5 years) and only a few students exceeded the standard period of the physics study (3.3\%).  The students attended 5.2 courses on average, 3.4 of which were physics courses (66\%). The non-physics courses consisted of mathematical courses, educational courses, soft skills, and other. More than one third of the students reported spending more than 40 hours studying per week, and about one out of five students reported investing less 20 hours (Figure \ref{fig:sample}). Half of the students claim that they invested more time for studying during the COVID-19 summer term 2020 than in the previous semester.

Concerning the technical resources, almost all students reported to have permanent access to a PC (95\%) and to a fast and stable internet connection (81\% agreement). However, only about half of the students had access to a printer (54\%).

\subsection{Data analysis}
\subsubsection{Construction of variables}
In order to reduce and condense the amount of data (max. 246 entries per student), common factors (variables) have been defined; the variables are a combination of items that aim at measuring the same construct.  To check whether items actually belong together as intended, i.e., whether they form a common factor, correlations between the items are used. This results in a small number of artificial variables that account for the most covariance in the observed variables, facilitating data interpretation. The single responses to individual items are then aggregated within the variables per student, yielding a distribution of scores for the population (e.g., see Fig. \ref{fig:appendix_normal}).

The a priori construction of items that fit together may not always hold empirically, that is, items may turn out to be inappropriate (measuring something different) or a scale might split into more than one component. This construction process uses factor analysis (principal component analysis) and content-interpretation. This process and results are not reported here; all items that belong to the scales and all excluded items are listed in the appendix. 

\textit{Scale reliability.}  The internal consistency of a component is measured by the reliability coefficient Cronbach’s $\alpha$. The measure indicates how well the items “fit together”, that is, it examines whether or not a test is constructed of parallel items that address the same construct. Cronbach’s $\alpha$ ranges from 0--1 and values above 0.7 are considered reliable for group measurement. 

\subsubsection{Correlation analysis and between-groups comparisons}
For determining the relationships between different variables, Pearson's correlation coefficient is used. It measures the linear correlation between two variables X and Y and ranges from -1 to 1. A value of +1 means total positive linear correlation, 0 is no linear correlation, and -1 is total negative linear correlation. 
To analyze the differences among group means in a sample, analysis of variance (ANOVA) is used. ANOVA provides a statistical test of whether two or more means of a dependent variable (as defined by the components) are equal among different groups, and therefore generalizes the t-test beyond two means.

\section{Results}

\begin{table*}[t!]
\centering
\caption{Description and psychometric characteristics of the scales that were used for all students}
\label{tab:descr}
\begin{tabular}{p{3cm}cp{5cm}cc}
\hline \hline

Scale	& ~~~\# items~~~&Sample item	&~~~~~$\alpha$~~~~~ &	Mean $\pm$ standard deviation (\%) \\ \hline \addlinespace[3pt]
self-organization abilities in general	& 5 &	In my studies, I am self-disciplined and I find it easy to set aside reading and homework time. &	0.76	&2.97 $\pm$ 0.57
(65.6 ± 25.5)\%\\\addlinespace[6pt]
self-organization abilities during COVID-19 semester	&6&	Not being at university hinders me from studying$^*$. &	0.77	&2.41 $\pm$ 0.67
(47.1 $\pm$ 22.4)\% \\\addlinespace[6pt]
Environment	&2	&I have a quiet space where I can participate in video conferences unhindered.	&0.90	&3.11 $\pm$ 0.76
(70.4 $\pm$ 19.1)\%\\\addlinespace[6pt]
Attitudes towards online learning	&6&	On-campus instruction helps me to understand the physics concepts better than in online courses.	&0.75&	2.84 $\pm$ 0.80
(61.5 $\pm$ 26.5)\% \\\addlinespace[6pt]
Communication	&8&	It is easy for me to establish contact with other students during Covid-19 pandemic.	&0.88&	2.19 $\pm$ 0.57
(39.5 $\pm$ 18.9)\% \\\addlinespace[6pt]
Learning achievement	&7&	I am certain that I will complete the online physics courses with good grades.	&0.82&	2.56 $\pm$ 0.66
(51.9 $\pm$ 22.6)\% \\ \hline \hline
\multicolumn{5}{l}{$^*$Negative statements were reversed for the analysis.}

\end{tabular}

\end{table*}

\subsection{Scale descriptives and reliabilities}
\begin{figure*}[t!]
\centering
\includegraphics[width=.92\linewidth]{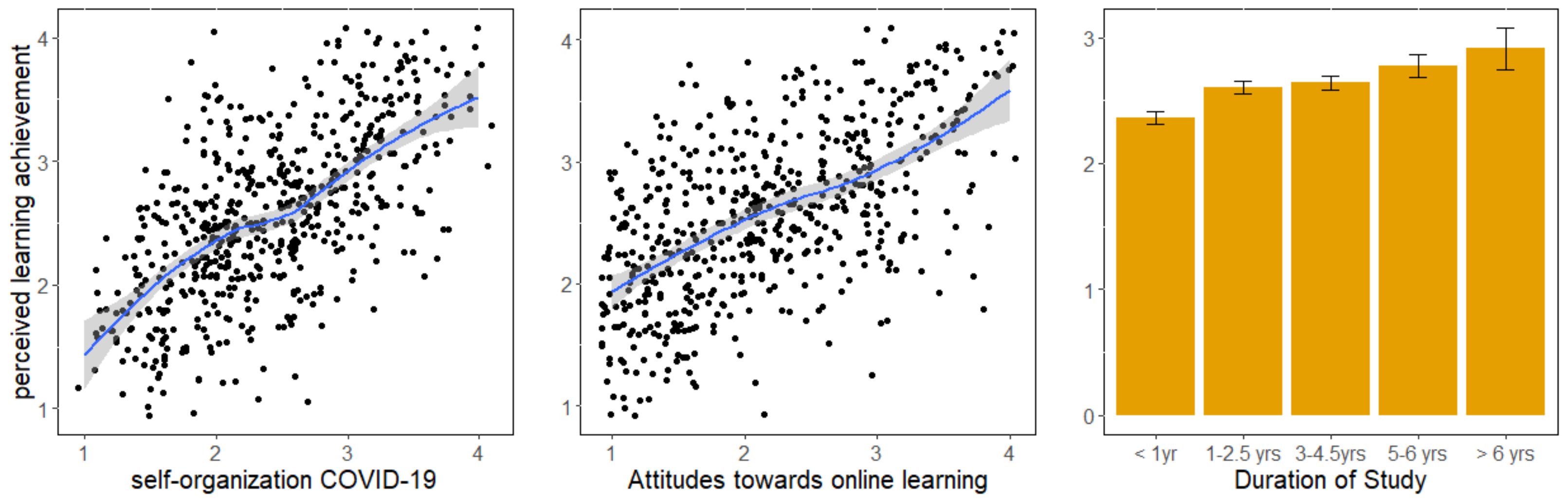} 
\caption{Scatter plots of learning achievement vs. self-organization abilities during COVID-19 pandemic (left) and student attitudes towards online learning (center), including a smoothed fit curve with confidence region. A small jitter was added to the data points to avoid overlapping. High values represent high success / abilities / preferences. Right: Learning achievement (perceived by the students) vs. duration of study. The error bar represents the standard error of the mean.}
\label{fig:correlmain}
\end{figure*}

As described in the method section, we constructed a multi-scale questionnaire to assess several aspects of online teaching and learning. Table \ref{tab:descr}  shows the characteristics of the scales. All items were assessed with a 4 point Likert-type scale, thus the means range between 1 and 4 points, where high values correspond to high abilities / preferences. For better interpretation, the means were also linearly transformed into percentages. Note that all scales were positively checked to be one-dimensional using explorative factor analysis. According to the histogram plots (Fig. \ref{fig:appendix_normal} in the appendix), we can also assume normal distributions for most of the scales.

In terms of absolute values, the ability to establish communication with peers and lecturers received the lowest scores, followed by the self-organization abilities of the students in this specific situation. The assessment of the learning achievement in this semester was normally distributed with a mean score of about 52\%. The personal environment of the students, which was assessed in terms of the ability to work without noise or distributions, received high scores.

\subsection{Correlation analysis and group comparisons (RQ1 and RQ2)}
To investigate the relationships between the scores, a correlation analysis was performed. Tab. \ref{tab:correl} shows the correlation coefficients for each pair of variables. To account for multiple testing, Bonferroni-correction was used to determine the level of significance ($p=0.05/15=0.003$). We observe that students' perceived learning success is positively correlated with high abilities of self-organization in general and in this specific situation, with the access to a study-friendly environment and to the ability to communicate with peers and lecturers. It is negatively correlated with students' attitudes towards online learning. In other words, students who have a positive attitude towards online learning also expect a higher learning achievement. The strongest relationships are displayed in Figure \ref{fig:correlmain}. Furthermore, it can be observed that the preference for face to face teaching is negatively correlated with all other scales (expect self-organization abilities in general), and all other scales are positively related to each other.

To investigate how different groups of students responded to the different types of questions, a multivariate analysis of variance (MANOVA) with the duration of studies (DoS, cf. Table \ref{Tab:sample}) as independent (group) variable was performed.  MANOVA examines differences on several dependent variables simultaneously and is used in this case because the dependent variables are intercorrelated. The results show that DoS had a significant multivariate influence across all variables, Wilks $\lambda = 0.87$, $F(30, 2250) = 2.76$, $p < 10^{-6}$. Given the significant multivariate effect, ANOVAs are then performed for each dependent variable using Bonferroni-correction to determine significance levels. Duration of studies had a significant impact on perceived learning achievement [$F(5, 567)=13.1$, $p<10^{-4}$], see Fig. \ref{fig:correlmain} (right), but on no other variable. In the next step, a post hoc analysis determines which levels of the dependent variable (DoS) differ from the others in terms of perceived learning success. Pairwise comparisons revealed that students who study physics for less than 1 year obtain significant lower scores than all other groups of students ($p=0.01$ to $p=0.001$). There is no statistical significant difference between any other student groups regarding this variable.  

\begin{table}[h!]
\centering
\caption{Correlation analysis. Only significant correlations (Pearson’s $r$; $p < 0.003$) are presented.}
\label{tab:correl}
\begin{tabular}{lccccc}
\hline \hline 
	& \multicolumn{5}{c}{Scales}							\\
	&	1	&	2	&	3	&	4	&	5	\\ \hline \addlinespace[4pt]
(6) Learning achievement	&	0.33	&	0.63	&	0.32	&	-0.58	&	0.50	\\
(1) self-organization general	&	1	&	0.37	&	0.17	&	n.s.	&	0.15	\\
(2) self-organization COVID-19	&	…	&	1	&	0.33	&	-0.62	&	0.55	\\
(3) Environment	&	…	&	…	&	1	&	-0.19	&	0.29	\\
(4) Attitudes towards onl. learn. 	&	…	&	…	&	…	&	1	&	-0.53	\\
(5) Communication	&	…	&	…	&	…	&	…	&	1	\\
\hline \hline 

\end{tabular}
\end{table}

\subsection{Recitations (RQ 3)}
\subsubsection{General information and recitation organization}
There are 401 data sets from students who attended recitations, 287 of which were related to introductory physics courses with large audience ($>$30 students), 91 of which were related to special courses with few students (e.g. special lectures in physics master studies), and 42 others (mathematical physics or special courses for prospective physics teachers). Note that some students took part in more than one course. In almost all cases (97\%), exercise sheets were provided and discussed in weekly online meetings with a lecturer (89\%). The exercise sheets were compulsory in 76\% of the cases. 
Students reported high engagement when solving the problem sheets; 82\% of the students attempted to solve more than 60\% of the tasks and 58\% dealt with more than 80\% of the tasks. The students worked on the tasks in groups in 50\% of the cases; due to the pandemic lockdown, they did not meet in person (only 4\% did) but rather organized their meetings mostly via web conference or used messengers for communication.
\subsubsection{Recitation formats: What was used and what would be optimal?} \label{sec:tutforms}
\begin{figure}
\centering

\includegraphics[width=\linewidth]{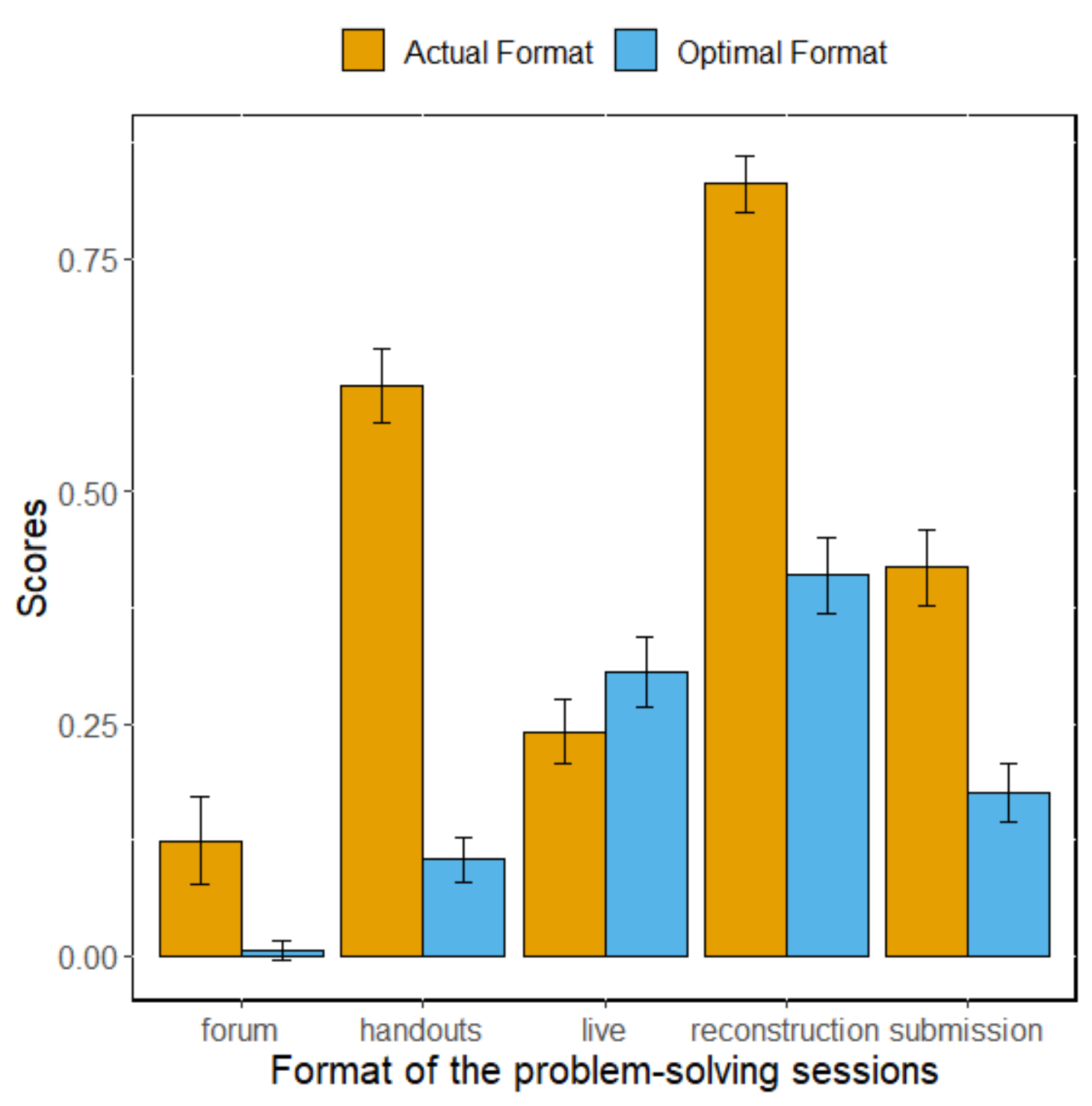}
\caption{Formats of the online recitations that students actually experienced during the COVID-19 summer term and judgments about optimal recitation}
\label{fig:tuts}
\end{figure}

Based on the student interviews, different online formats of the recitations were identified which could be assessed in terms of the following aspects (true or false): 
\begin{itemize}
\item (submission) The students' solutions were submitted, corrected by the tutor and discussed in an online meeting
\item  (reconstruction) The solutions were reconstructed during a live online session in real time
\item  (live) Exercise sheets were solved online live and discussed in groups 
\item  (handouts) The solutions to the exercise sheets were handed out to the students as text or as a video (worked-out solutions)
\item (forum) Forums were used to discuss exercise sheets without time constraints.
\end{itemize}
Students marked what formats they actually experienced (more than 1 option could be chosen) and what format they consider optimal for future online recitations (single choice). Figure \ref{fig:tuts} shows the fraction of total answers for each category. Due to the different assessment types (multiple choice vs. single choice), statistical comparisons are inappropriate. However, comparing the relative counts within each data series, it can be observed that students are less interested in handouts of solutions, but wish to work together live on exercise sheets. In line with that result, we observe that forums are perceived to be ineffective for the recitations.

\subsubsection{Perceived effectiveness of recitations}
For evaluating the perceived effectiveness of the online recitations, we assessed the student’s agreement concerning seven questions on a 4 point Likert-type scale, e.g. whether they felt well supervised concerning their questions during e-learning or whether the recitations were helpful to foster their understanding. The scores are normally distributed (mean = 2.20, corresponding to 40.2\%, see Fig. \ref{fig:appendix_normal} left) and the reliability of this scale is $\alpha = 0.82$. It is worth noting that the perceived effectiveness of the recitations is highly correlated with the perceived overall learning achievement  ($r = 0.66$, $p<0.01$).

To investigate whether different groups of students or different course formats had an impact on the students’ perceived effectiveness of the online recitations, ANOVAs were performed. First, a significant between-subjects effect was found for the duration of studies [$F (4, 389) = 5.04, p= 0.001$] and Bonferroni-corrected post-hoc comparisons revealed that students who spent more than 6 years studying physics perceived the recitations as more effective than every other group of students (mean score: 2.95, corresponding to 65\%; $p<0.01$). Second, groups of students were defined based on their experienced format of the recitations (see Sect. \ref{sec:tutforms}). It was observed that students who sent their own solution to be rated ($N=168$) assessed the effectiveness of the recitations significantly higher than students who did not ($N=227$), $F(1, 393) = 6.74, p=0.01$. There were no significant differences concerning other group splits.

\subsection{The physics laboratories (RQ4)}
\subsubsection{General information and lab organization}
220 physics students were enrolled in physics laboratories that was adapted to an online lab in most cases (82\%). In few cases (10\%), students could visit the laboratories of the university or a mixture of live and online labs were used (8\%). For the online laboratories, the experimental work was enabled either (a) using simulations and evaluating simulated data / gathering data from the simulation, (b) using videos of real experiments and extracting data from the video or (c) working with data gathered by someone else. There were some single statements (less than 5\%) about other formats, e.g. replacing the labs by student talks or using smartphones for data collection at home that were not further considered. Hence, three different student groups according to the data source can be defined (simulated data, real data gathered by students themselves, real data gathered by someone else).

\subsubsection{Effectiveness of the physics laboratories}
To evaluate the perceived effectiveness of the physics laboratories, eight questions were included into the questionnaire. One set of questions deals with the experimental skills that were acquired during the lab course (5 questions, $\alpha=0.75$, sample item “I gained less experimental skills due to the modified course format”) and another set of questions asked about reinforcing content (3 questions, $\alpha=0.63$, sample item “The modified physics lab helped me to better understand the physics concepts behind the experiments”). The distribution of both scales is depicted in Fig. \ref{fig:appendix_normal}.

\begin{figure}[h!]
\centering

\includegraphics[width=\linewidth]{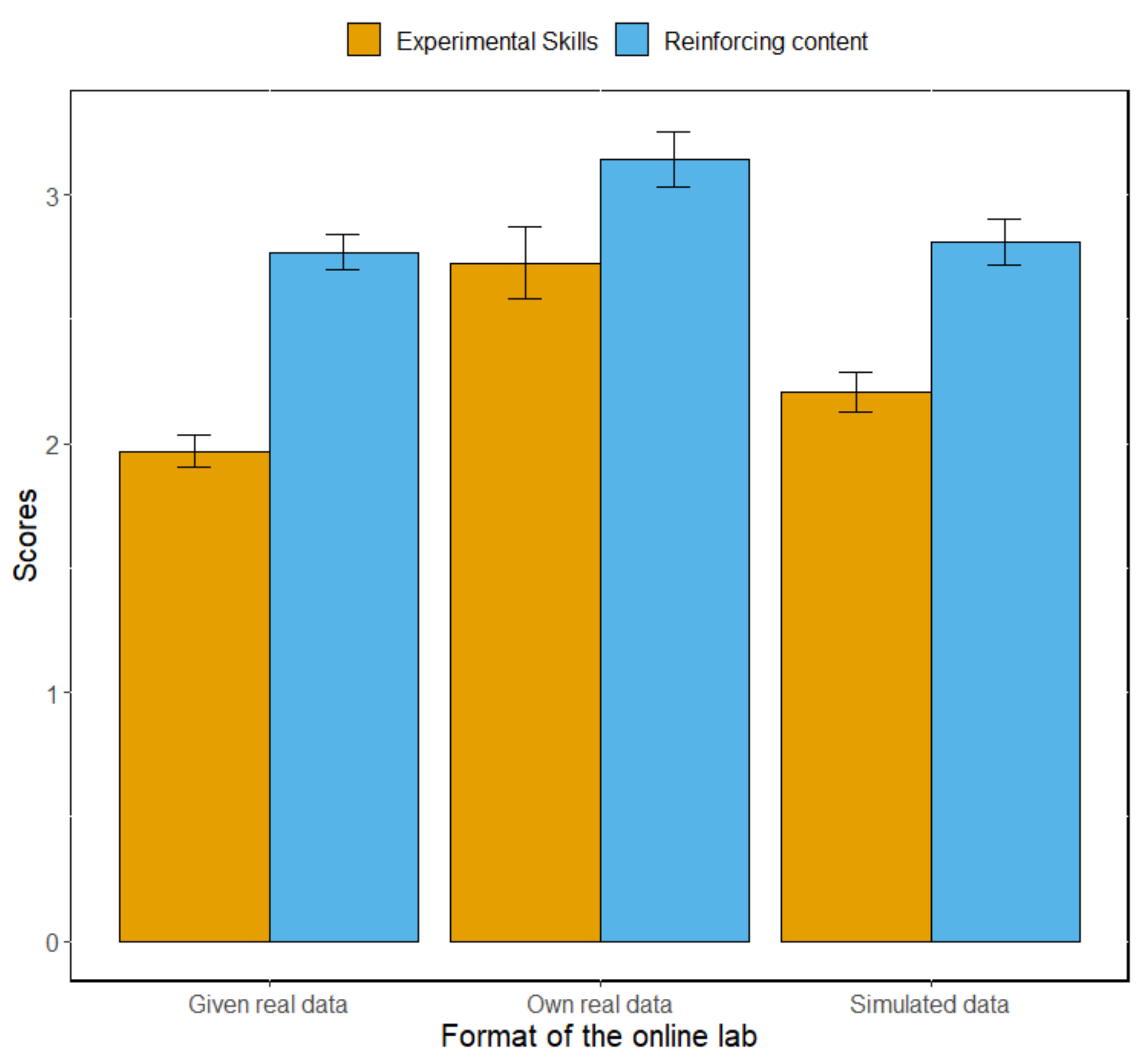}
\caption{Evaluation of the physics lab with respect to different types of data}
\label{fig:lab}
\end{figure}

In order to find out whether different groups of students perceived the two aspects of the physics laboratories differently, ANOVAs were conducted. First, the duration of the study had a significant impact on students’ perceived acquisition of experimental skills, $F(3, 204)=4.41, p<0.01$. Students in their first academic year perceived the labs as less effective concerning experimental skills compared to students in their second year. There was no such difference concerning the reinforcement of content. 
Second, the type of data that was analyzed by the students had a significant impact on both scales, the acquisition of experimental skills [$F(2, 198) = 14.2, p<0.001)$] and the reinforcement of content [$F(2, 216)=3.50, p=0.03]$. Bonferroni-corrected post-hoc comparisons revealed that students assessed significant higher scores when they gathered the data by themselves (e.g. from a video) as compared to simulated data or data that was handed out, see Fig. \ref{fig:lab}. 

\section{Discussion}
\subsection{Factors that are correlated with subjective learning achievement during COVID-19 pandemic}
From the measurement perspective, all questionnaires that have been used show a good reliability, and the students' scores were reasonably normally distributed. The perceived learning achievement was positively correlated with students' self-organization abilities ($r=0.33$), particularly in the COVID-19 pandemic ($r=0.63$). While researchers have pointed out the benefits of metacognitive strategies such as self-regulation of learning on problem-solving skills before \cite{Rickey2000, Schraw06}, the correlation  between self-organization skills during the COVID-19 pandemic and learning achievement ($r=0.63$) was surprisingly strong. Even though no causal relationship was proven, we encourage physics faculties to offer special courses for promoting self-regulated learning skills, e.g. including time management training, encouraging note-taking, setting pace, etc. The need for that is demonstrated by the rather low mean value of the scale (47.1\%, cf. Fig. \ref{fig:appendix_normal} and Tab. \ref{tab:descr}). During the interviews that have been conducted beforehand, many students found it difficult to structure their daily routines and keep up with their usual learning pace. Supporting students with strategies how to improve their self-organization might presumably also improve students' learning achievement.  

 Moreover, we also found a high correlation between communication and learning achievement. One explanation might be that high-performing students are better socialized and therefore find  communication opportunities more easily than lower-performing students. Or vice versa, good communication is a prerequisite for successful study, especially in physics, where complex phenomena and problems are studied and solved in pairs and groups. Interactions with lecturers and  peers can help  to create an atmosphere of commitment to understanding \cite{Dufresne}. When engagement with lecturers and other interested learners is hindered due to the distance, weaker students are potentially left behind. A constructivistic perspective advocates active learning formats, where students are engaged in writing, talking, describing, explaining, and reflecting---these processes that require careful thinking and planning when switching from the lecture hall to online lectures. To promote communication between the lecturer and the students,  the  instructor  not only presents material but also asks questions, uses interactive quizzes, and stops periodically to encourage group discussions; just what makes a good face-to-face lecture as well \cite{Mazur93, Sokoloff94, VanHeuvelen91}. To establish communication between the students, networking services can be installed where peers can meet and form learning groups. During the online classes, break-out rooms can create an atmosphere in which a small group of students can exchange ideas without permanent (online) presence of an instructor. Keeping the groups small can foster collaborations.
 
 For the attitudes towards online learning, we found a negative correlation to the perceived learning achievement. A positive attitude towards distance learning was found to be positively correlated to subjective and objective academic learning achievement in other work \cite{Uzun}, and we also found that this attitude is correlated with self-organization skills and communication. Besides improving communication and self-organization skills as suggested above, we further  recommend that instructors make the students aware of the positive aspects of distance learning, i.e. permanent access to video-taped lectures, easier implementation of simulations and videos, saving traveling time, etc.  

\subsection{First-year students rate learning achievement the lowest}
The analysis of different student groups based on their duration of studies revealed that the first-year students reported significant lower subjective learning achievement than all other groups of students. Possible explanations for this may be that (1) the first-year students have had the least contact with other students at university so far and therefore communication suffers, (2) self-organization skills of experienced students are more pronounced, (3) the first-year students have had less contact with online teaching so far and therefore their attitude is also worse, or (4) that they are  more cautious in their performance assessment, regardless of the online teaching situation, because they have had less experience of success than older students (who are still studying). There was no difference concerning communication, self-organization skills, and students attitudes towards online teaching between the different age-group of students, hence (1)--(3) cannot be supported by the data. However, our data set contained no students who started their study in spring 2020, so all students in the lowest duration group were in their second academic semester. Extrapolating the trend given in Fig. \ref{fig:correlmain} (c), we can assume an even larger gap to the freshmen that start in the next terms. When faculties start to offer on campus teaching again with limited capacities, we therefore suggest focusing on freshmen students.

In line with this result, we found that students who spent 6 years or more studying physics rate the effectiveness of the recitations higher than the younger students and the first-year students rated the perceived acquisition of experimental skills the lowest.  Note that the sample size is different for each comparison (since not every student took part in recitations or laboratories).

\subsection{Recitations and tutors' feedback on solutions are important for subjective learning achievement}
For the recitations, a high correlation between perceived effectiveness of the recitations and subjective learning achievement was found. The exercise sheets (tasks) to be worked on in self-study form a link between reception of the course contents during lecture and active work on the associated problems. Given the high degree of self-activity of the students, the recitations are often regarded as the decisive basis for actual learning, contributing significantly to the knowledge construction of the students and are essential for the understanding of physics \cite{finkelstein, Spike}. Students were asked how the recitations were designed during the COVID-19 pandemic and what format they would prefer. A major discrepancy between actual and optimal format is present for the handout of solutions. Students prefer live discussions and reconstructions of problem solutions over the distribution of solutions. Furthermore, the students who received graded feedback on their own solutions perceived the recitations as more effective than students who received no feedback. Both results are plausible from an educational perspective; first,  group discussions, corporate problem-solving activities and active exchange are crucial parts of face-to-face recitations and cannot be replaced by the handout of solutions that requires retracing the solution on their own. Second, it is known that receiving cognitive feedback can be associated with increased performance \cite{Duncan07}, and especially weaker students benefit from feedback comments---regardless of the actual quality of the feedback \cite{Gielen10}. Feedback on own solutions make errors visible, can evoke new ideas how to approach problems, and help to identify weaknesses that otherwise remain undetected.

\subsection{The physics laboratories were considered most successful when own data was collected}
220 students took part in an online physics lab course that was mandatory for the participants. We did not evaluate the intended learning goals of the physics laboratories but decided to ask about the reinforcement of content and building of experimental skills. These two goals are present in the  research literature around lab courses and most often, the courses focus on skills or on both aspects \cite{Wilcox17}. The items asked, for example, about conceptual  mastery  of  the  subject  matter (reinforcing content) or assessment of students' ability to make measurements and collect accurate data (experimental skills), respectively. 

The students most often reported performing measurements through videos of experiments (both videos recorded by the instructor or publicly available), to use simulations to generate data or just to analyze given data sets with no engagement to experimental equipment. Some students reported watching an instructor doing the experiment, having a description and a picture of the experiment, or controlling physical equipment remotely. Based on the variety of responses, we identified three categories to characterize the data students analyzed: (1) students were supported with real data taken from someone else, (2) students used simulated data taken by themselves, and (3) students gathered data by themselves from a real experiment (that was videographed, remotely controlled or actually performed). Earlier studies suggested that it was possible to use secondhand data for the purpose of evaluation and interpretation without significant distortions of epistemic learning processes \cite{Priemer}. 

In contrast, our results  show that students perceive higher learning success when students gathered data on their own, regarding both, reinforcing the physics content and acquisition of experimental skills. While Priemer \textit{et al.} (2020) paid attention that students got enough information on how the data were generated, we did not ask for this information. In cases where using secondhand data is preferred, we therefore encourage lecturers to provide this information. Furthermore, in the study by Priemer \textit{et al.}, students engagement with the experiment was controlled among the students (working with firsthand or secondhand data). In our case, working with secondhand data might have been connected with lower engagement since data evaluation does not necessarily require dealing with the  apparatus. In contrast, gathering own data does.  

\section{Conclusion} 
In this descriptive study, more than 500 students from five European universities were assessed to obtain information how studying physics during the COVID-19 pandemic was experienced in Spring 2020. For this purpose, a multidimensional questionnaire was developed that turned out to have satisfactory psychometric properties, nevertheless revisions during future iterations will follow.  Here, we reported results about online problem-solving sessions, the online physics laboratories, and the factors influencing subjective learning outcomes. Even though being descriptive in nature (and therefore lacking controlled experimental manipulations), we derived several suggestions from the study for future physics courses in which online learning will certainly play an important role. We summarize our main findings: First, since self-organization skills are strongly correlated with subjective learning outcomes, physics faculties may offer (or import) special courses (or workshops) to foster these skills among students.  Second, since there was also  a strong correlation between subjective learning achievement and communication, we recommend installing networking services, using break-out rooms and keeping the groups small to foster communication between peers;  to promote communication between the lecturer and the students,  the  instructor can make use of interactive learning tools that might even be easier to implement in an online setting compared to the physical lecture hall. Third, first-year students rated the lowest learning achievement during the Spring term 2020. We conclude that all these interventions are especially important for the younger students who just started their academic career. When capacities for face-to-face teaching can be used to a limited extent, focusing on the freshmen cohort is recommended \cite{Unigoe}.  

For the problem-solving sessions (recitations), graded feedback followed by online discussions turned out to be more effective (in the students' perspective) than handed solutions. Making the solution process visible (reconstruction) requires technical resources, e.g. tablet computer with pen input. For the physics laboratories, students value the possibility of gathering own data. This can be accomplished from the distance via remote controlled labs, or (more easily) with videos or simulations from experiments. When secondhand data is used, we advice to clarify their origin, e.g. with a detailed description of the measurement 
including photographs of the experimental setup. 

This study has some limitations. From the methodological perspective, we relied purely on self-assessments and subjective estimations of learning and attitudes. Even though the results are consistent with educational notions, future work could also investigate exam results or other hard indicators of learning for a more objective data basis. Furthermore, the sample represents a selection of universities that were accessible to the researchers. A huger distribution of the questionnaire, involving more countries and different institutions would increase the generality of the results. Last, the questionnaire was administered only once as a mid- to end-term evaluation during the Spring 2020. When students get more accustomed to online learning or hybrid models of learning, some variables (self-organization or communication) might improve incidentally.

The result presented here might be helpful for other faculties for planning online teaching and learning in general, and during the next terms. The close cooperation with colleagues from the physics departments  has proven to be very successful, especially in the implementation of the questions, distribution and reporting of results during faculty meetings. On a meta-perspective, the article therefore also shows how education research can bridge the (sometimes existent) gap between physics education and the faculty of physics.  It would be desirable if similar work has more space in the future and also at other institutions.

\appendix
\section*{Appendix}
\begin{figure*}[h!]
\includegraphics[width=.9\linewidth]{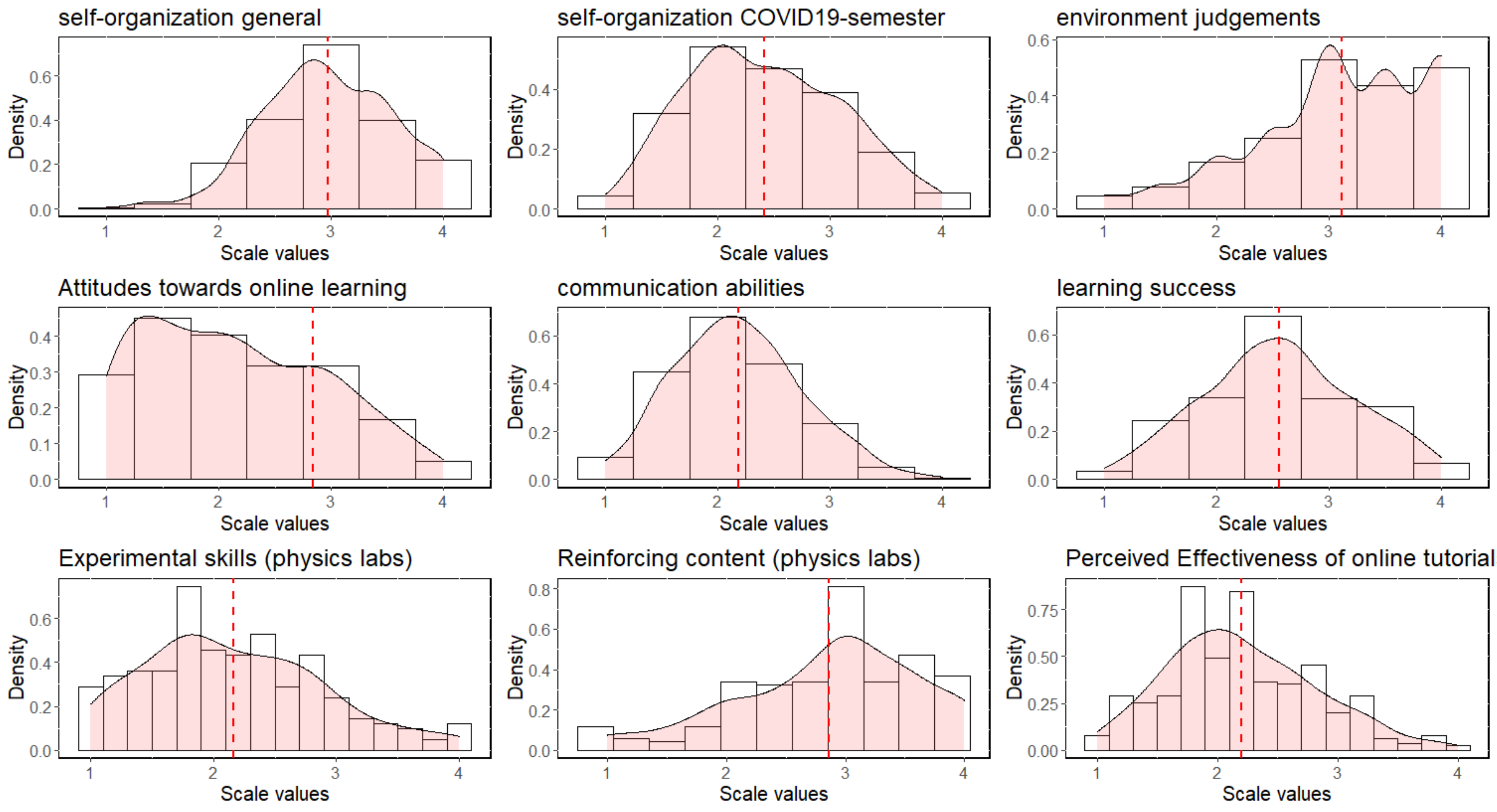}
\caption{Distribution of student’s self-assessments of the six scales described in Tab. \ref{tab:descr} and of the special physics courses (recitations in terms of effectiveness and the labs in terms of achieving experiment skills and conceptual understanding)}
\label{fig:appendix_normal}
\end{figure*}

The distributions of the students' scores are depicted in Fig. \ref{fig:appendix_normal}. 
\ \\
\ \\
\textbf{Questionnaire items:} \\
\ \\
\textit{a) self-organization abilities in general}
\begin{itemize}
\item In my studies,  I am self-disciplined and I find it easy to set aside reading and homework time.	
\item I am able to manage my study time effectively and complete assignments on time.	
\item In my studies,  I set goals and have high degree of initiative.	
\item When it comes to learning and studying,  I am an self-directed (independent) person.	
\item I plan out my week’s work in advance,  either on paper or in my head.
\end{itemize}
\textit{b) self-organization during COVID-19 semester:}
\begin{itemize}
\item Not being at university hinders me from studying.
\item I am not able to organize my time in the during e-learning effectively.
\item I lack the daily routine due to absence of classes at university.
\item I find it difficult to get up in the morning without having a scheduled class.
\item I manage to complete the assignments for the online courses.
\item I am more systematic and organized during Covid-19 pandemic than usual.
\end{itemize}
\textit{c) environment}
\begin{itemize}
\item I have a quiet space where I can participate in video conferences unhindered.	
\item I am disturbed by others during times when I need to study for my courses.
\end{itemize}
\textit{d) attitudes towards online learning}
\begin{itemize}
\item In general, I prefer the online physics courses over classroom physics courses.		
\item For the same type of course, my classroom learning experience in physics is better than my online learning experience 
\item On-campus  instruction helps me to understand the physics concepts better than in online courses.	
\item I feel more passive in online physics courses than in the standard classroom courses on campus.
\item The same instructor encourages more discussion in standard classes than in online classes.	
\item Based on my experience this semester,  I look forward  to have more online physics courses even when the standard courses at university will be available.	

\end{itemize}
\textit{e) communication}
\begin{itemize}
\item I miss having contact with other students due to Covid-19 pandemic.		
\item The communication with the instructors works fine during COVID-19 pandemic.	
\item The communication with peers works fine during Covid-19 pandemic.
\item At the moment, I miss discussions about physics topics with other students at university.		
\item I hesitate to ask my peers for help in e-learning environment.		
\item I meet online or on telephone with other students to learn together.	
\item It is easy for me to establish contact with other students during Covid-19 pandemic.	
\item It is hard for me to discuss with other students when I don’t see their faces.	

\end{itemize}
\textit{f) overall learning achievement} 
\begin{itemize}
\item I believe I can perform well in an online self-paced class.
\item I am confident that I can learn without the permanent presence of an instructor to assist me.
\item I am confident that I can manage all the activities that the online physics classes require.	
\item I look anxiously at the upcoming exams in physics.		
\item I am certain that I will complete the online physics classes courses with good grades.	
\item I am afraid that I will not be able to perform as I am used to given the unfamiliar situation.		
\item Studying during the COVID-19 pandemic has positive impact on my study success.

\end{itemize}
\ \\
\textbf{Lab courses}
\ \\
\textit{General:}
\begin{itemize}
\item Did you attend a physics lab course this semester?	
\item How was the lab work conducted?For example: I conducted experiments by myself (real interactions with experimental equipment), I performed simulations or virtual experiments, I performed remote experiments, I received video records of experiments or other.
\item Which type of data did you analyze?
\end{itemize}
\textit{Reinforcing content:}
\begin{itemize}
\item Despite major changes in the course structure,  the lab course improved my knowledge.
\item The online physics lab helped me to better understand the physics concepts behind the experiments.
\item Online discussions about the labs with the experiment supervisor or with peers reinforced my physics knowledge. 

\end{itemize}
\textit{Experimental skills:}
\begin{itemize}
\item The limited hands-on activities had negative impact on my learning		
\item The online lab course in this semester had a similar effect on my experimental skills as the standard lab course.
\item Not having gathered the measurement data myself complicates data analysis and discussion		
\item Missing the hands-on experience,  I am concerned that I cannot handle experimental equipment in the next lab/in the future.		
\item I feel that I gained less experimental skills due to the modified course format.

\end{itemize}

\ \\
\textbf{Problem-solving sessions / recitations}\ \\
\textit{General}
\begin{itemize}
\item Did you attend online problem-solving sessions during the semester?
\item To which type of course did the sessions belong to? E.g. Introductory physics lecture with a relatively large audience, special course in physics with few students or other?
\item Were exercise sheets provided?	Were exercise sheets compulsory?
\item What fraction of the exercise sheets did you complete?	
\item Did you work alone or with peers on the tasks?
\item If you solved the homework problems in a study group,  how did you organize the study groups?
\item Did you meet online to discuss the solutions to the tasks?	
\item If online discussions of the problem sheets existed,  what fraction did you attend?
\end{itemize}
\textit{Structure}
\begin{itemize}
\item How was the problem-solving course structured? Click all that apply and/or describe the structure yourself.
\begin{itemize}
\item The students' solutions were submitted, corrected by the tutor and discussed in an online meeting
\item  The solutions were reconstructed during a live online session in real time
\item  Exercise sheets were solved online live and discussed in groups 
\item  The solutions to the exercise sheets were handed out to the students as text or as a video (worked-out solutions)
\item  Forums were used to discuss exercise sheets without time constraints.
\end{itemize}
\item Which format (one of the above) would you consider  best in terms of fostering your learning achievement? Think about what format you would like to see in the future.
\end{itemize}
\textit{Perceived effectiveness of the problem-solving sessions} \ \\
\begin{itemize}
\item  The online problem-solving sessions in this semester had a similar effect on understanding physical concepts as the regular exercises.	
\item The online format led to less discussion about the individual solution steps.	
\item In regular recitations a better exchange between students and teachers is achieved.
\item  When presenting solutions online, it is easier for me to understand the individual steps and approaches.
\item During the online sessions I felt well supervised concerning my questions 	
\item Lacking handwriting opportunities in online sessions makes discussions about the tasks less effective		\item Working on exercise sheets in a study group was more difficult than in normal semesters.
\end{itemize}

\pagebreak


\begin{thebibliography}{99}
\bibitem{Unigoe} Indeed, the Faculty of Physics of the University of Goettingen prioritizes the first-year students and live problem-solving sessions: In the following winter term, the large lecture halls will be used for problem-solving sessions (instead of giving lectures) for the first-year courses. The occupancy rate will be 20\% to prevent the spread of the Corona-virus.
\end{thebibliography}
\end{document}